\documentclass[superscriptaddress,prl,twocolumn]{revtex4}
\usepackage{graphicx}
\usepackage{float}
\usepackage{color}
\usepackage{amsmath}
\usepackage{amssymb}
\usepackage{natbib}

\begin{document}

    \title{Refraction of fast Ne atoms in the attractive well of LiF(001) surface}
    \author{M. Debiossac}
	\author{P. Roncin}
    \author{A.G. Borisov}
    \affiliation{Institut des Sciences Mol\'{e}culaires d'Orsay (ISMO), UMR 8214 CNRS -   Universit\'{e} Paris-Sud, Universit\'{e} Paris-Saclay, b\^{a}t 520, Orsay, France}
    \date{\today}

    \begin{abstract}
Ne atoms with energies up to 3 keV are diffracted under grazing angles of incidence from a LiF(001) surface. For a small momentum component of the incident beam perpendicular to the surface, we observe an increase of the elastic rainbow angle together with a broadening of the inelastic scattering profile. We interpret these two effects as the refraction of the atomic wave in the attractive part of the surface potential. We use a fast, rigorous dynamical diffraction calculation to find a projectile-surface potential model that enables a quantitative reproduction of the experimental data for up to ten diffraction orders. This allows us to extract an attractive potential well depth of 10.4 meV.
Our results set a benchmark for more refined surface potential models which include the weak Van der Waals region, a long-standing challenge in the study of atom-surface interactions.

    \end{abstract}

    \maketitle
Refraction is a well known phenomenon occurring when light is deflected
at an interface between two transparent media of different refractive
indices. In the early days of quantum mechanics, refraction of matter-wave
was first observed with electrons \cite{Davisson1928,Kikuchi1928} and was
explained by considering the beam to be refracted by the inner potential
of the material \cite{Yamaguti1930}. Refraction has been also reported for
neutrons \cite{Klein_1983}. For molecular and atomic projectiles
scattered from surfaces at thermal energies \cite{Harris1982,Zhao_2011},
the refraction is produced by an attractive Van der Waals (VdW) part
of the projectile-surface  interaction potential \cite{Beeby1971}.
Thus, collision experiments between neutral atoms and surfaces represent an ideal platform for characterization of the physisorption region dominated by weak VdW polarisation forces  \cite{Farias1998,Hoinkes1980} which is of paramount importance
for various practical applications \cite{SelfAss,Andersson1998,Nijkamp2001,DeFeyter}.

\begin{figure} \includegraphics[width=0.75\linewidth,draft = false] {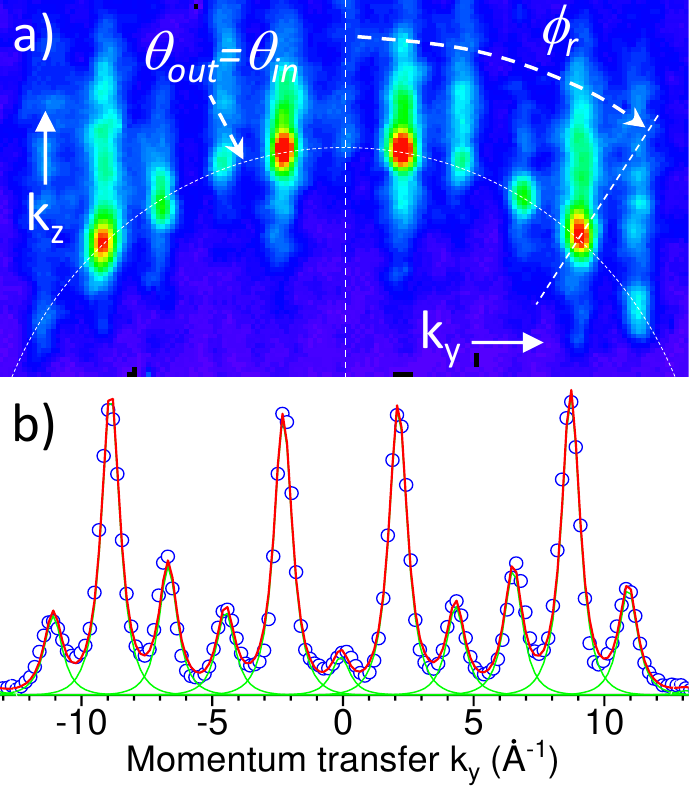}%{y13_h221.pdf}
	\caption{a) Diffraction pattern in the $(k_y,k_z)$ plane recorded for 500~eV Ne atoms scattered off LiF(001) surface along the  $\langle 110 \rangle$ direction. The Laue circle is shown with a dashed white line and corresponds to an grazing incidence angle $\theta=0.33^{\circ}$.  $\phi_r$ indicates the classical rainbow angle. b) Projected experimental intensity along the Laue circle (blue circles). The red line corresponds to a fit of the diffracted intensity using a single empirical Lorentzian line-shape for all peak (green line).  \label{fig:y13_h221_2D}}
\end{figure}

So far atom scattering studies of the attractive part of the surface potential
have been mostly performed using diffraction and refraction of the
projectile beams at thermal energies \cite{Hoinkes1980,Semerad_1987,Harris1982,Mantovani}.
Recent observation of the surface diffraction of fast atoms of keV
energies under grazing incidence (GIFAD or FAD) \cite{Schuller_2007,Rousseau_2007,Winter_PSS_2011}
offered yet unexplored possibilities of the surface analysis.
Grazing incidence conditions imply slow motion of the projectile
perpendicular to the surface during the fast motion parallel to it.
GIFAD thus combines the sensitivity of thermal atoms with the geometry
of reflection high energy electron diffraction, that allows one to
record the full diffraction pattern at once using an imaging detector (Fig.\ref{fig:y13_h221_2D}).
Fast atom diffraction has already proven to be sensitive to surface
rumpling at the pm scale~\cite{Schuller2010} and to the VdW potential \cite{Debiossac_PRB_2014,Debiossac_PRB_2016}, in particular with
the observation of bound state resonances~\cite{Debiossac_PRL_2014}.
Similar to the thermal atom diffraction, most of the GIFAD studies
have been performed with light (He, H, H$_2$) projectiles.
The diffraction experiments with heavier Ne atoms are challenging
because of enhanced sensitivity to defects \cite{Semerad_1987,Rieder1994,Taleb_2017,Sibener_2017,Anemone_JPCC_2017}.
Nevertheless, distinct quantum feature as supernumerary rainbows were measured ~\cite{Schueller_2008,Gravielle_2011} suggesting that high coherence elastic diffraction could be observed.

\begin{figure*}[ht]
	\centering
	\includegraphics[width=1.\textwidth]{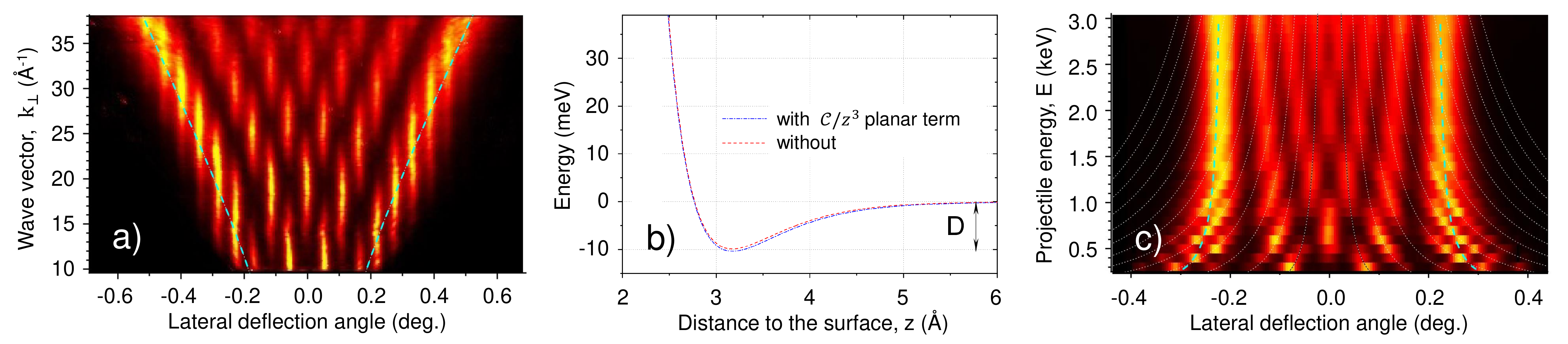}
	\caption{a) Experimental diffraction chart for $500$~eV Ne atoms scattered off LiF(001)  surface along the $\langle 110 \rangle$ direction. The grazing incidence angle $\theta$ is varied between $0.27^{\circ}$ and $0.95^{\circ}$. The intensity of the diffraction orders is shown as function of the lateral deflection angle, $\alpha$, and  $k_{\perp}$ the momentum component of the incident beam perpendicular to the surface. b) Theoretical planar potential $V(z)$ as a function of the distance $z$ from the LiF surface. The blue (red) line corresponds to the calculation with (without) an explicit planar attractive term. $D$ indicates the potential well depth.
%	For further details see the text.
	c) Same as a), but for the fixed incidence angle $\theta=0.42^{\circ}$ and total kinetic energy $E$ varied in $0.3 - 3$~keV range. The intensity of the diffraction orders is shown as function of the lateral deflection angle $\alpha$ and beam energy $E$.
The white dotted lines follow the position of the diffraction orders. The blue dashed line in panels a) and c) indicates the position of the rainbow angle calculated from Eq.~\eqref{eq:DeflAng} using $D=10$ meV.
		\label{fig:Charts}}
\end{figure*}

In this Letter we show well-resolved elastic diffraction of
fast Ne atoms on a LiF surface. We fully exploit the shorter
wavelength of Ne atoms so that many diffraction peaks
can be observed offering a surface probe with a resolution not obtainable with light projectiles \cite{Rieder_1984}. 
In particular, we reveal the refraction of the matter wave at the surface resulting in a shift of the elastic rainbow angle, and the broadening of the inelastic profile.
Rich diffraction patterns measured in our experiments provide further insights on the projectile-surface interaction from a detailed comparison with quantum simulation.
Our work, provides an important basis for a characterization of the physisorption region
for heavy atomic projectiles, which is a challenge both experimentally and theoretically.

A typical diffraction image is displayed in Fig.~\ref{fig:y13_h221_2D}. The LiF(001) surface
corresponds to the $(x,y)$-plane, and Ne projectiles are incident under grazing incidence angle
$\theta$ measured in the $(x,z)$ specular plane. The bright spots located on the Laue circle of energy conservation are due to elastic diffraction, whereas inelastic diffraction leads to the  elongated streaks \cite{Debiossac_Nim_2016,Roncin_PRB_2017}. In Fig.~\ref{fig:y13_h221_2D} the rainbow angle is denoted as $\phi_r$. Within the classical picture it corresponds to the largest angular deflection of the incident beam. In the quantum diffraction pattern, it is associated with intense diffraction spots at maximum lateral deflection. For the GIFAD scattering conditions, the
axial surface channeling approximation (ASCA) \cite{Rousseau_2007,Manson_PRB_2008,Zugarramurdi_2012,Diaz_2016b} provides a powerful
framework for the data analysis that we will use here. Within ASCA the fast motion along $x$-direction and the slow motion in the $(y,z)$-plane perpendicular to the surface are decoupled.  When the $x$-direction coincides with a surface crystal axis, the elastic diffraction pattern such as observed in Fig.~\ref{fig:y13_h221_2D},  is equivalent to that of a projectile with a wave-vector $k_\perp=k \sin \theta$ and energy $E_\perp$, incident at the potential $V_{2D}(y,z)$. Here, $k=\sqrt{2ME}$ is the total momentum, $E$ is the energy of the beam so that $E_\perp = E \sin^2 \theta$, $M$ is the projectile mass, and $V_{2D}(y,z)$ is the total projectile-surface interaction potential $V(x,y,z)$ averaged along $x$.

\begin{figure*}[ht]
	\centering
	\includegraphics[width=1\textwidth,draft = false] {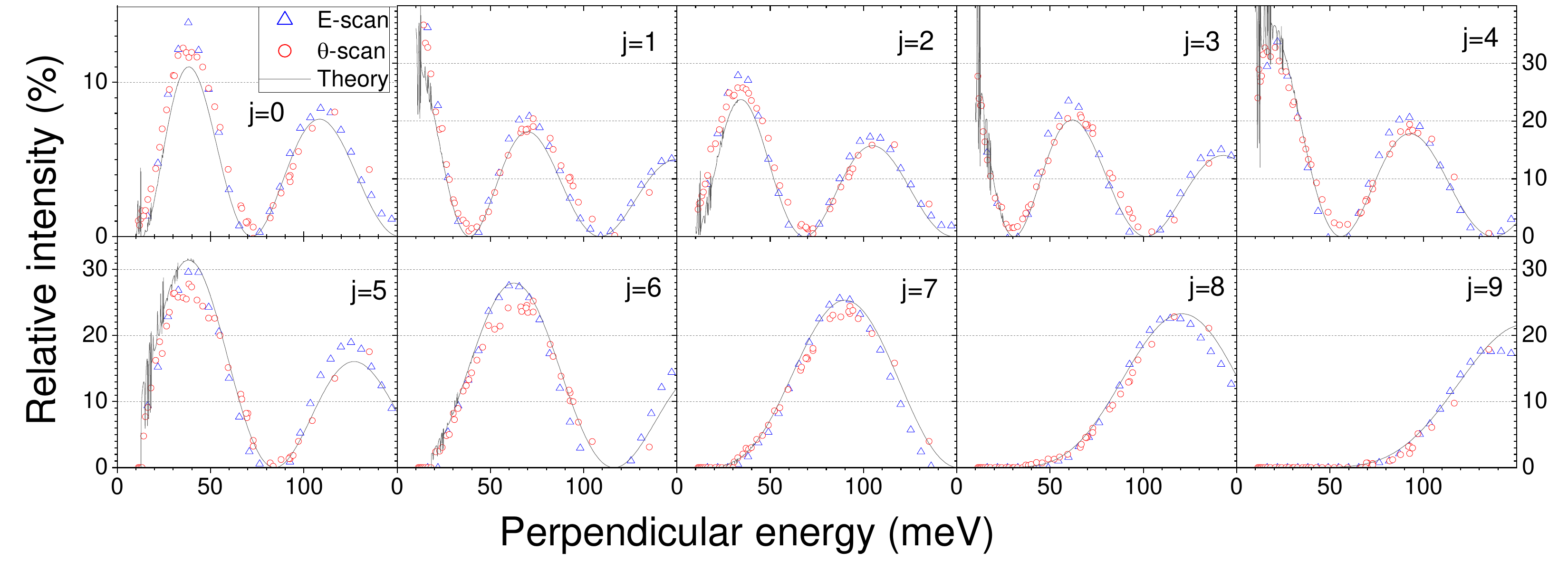} %off LiF(001) surface
	\caption{Rocking curves for Ne atoms scattered along the $\langle 110 \rangle$ direction. The intensity of the $j^{th}$ diffracted beam is shown as  function of $E_\perp$, the energy of the motion in the $(y,z)$-plane. For the degenerate $\pm j$ diffraction orders, the summed intensity is shown. Diffracted intensities measured during a scan of the projectile energy (blue triangles) or grazing incidence angle (red circles), the black line is for theoretical results. The very narrow peaks calculated at low energies reveal the bound state resonances.
		\label{fig:Exp_Ib}}
\end{figure*}

Prior to discussion of the experimental evidence of refraction effects we find
it useful to use the ASCA and to introduce a simple and intuitive model as a guide
to understand the main results of this work.

Consider the projectile-surface collision as a three step process
(i) acceleration by an attractive VdW interaction on the ingoing path;
(ii) collision with the surface leading to the deflection by the $V_{2D}(y,z)$ potential;
(iii) deceleration by the VdW interaction on the outgoing path.
Within this scenario, the largest lateral momentum resulting from the
scattering with the surface is
$k_{y}=\sqrt{k_{\perp}^2+2MD} ~ \sin \tilde{\phi}_r$.
Here, $D$ accounts for the acceleration on the ingoing path, and $\tilde{\phi}_r$
is the "true" rainbow angle given by the fastest
variation of $V_{2D}(y,z)$ with $y$ \cite{Schuller_EPL_2008}. The associated
maximum lateral deflection angle $\alpha$, and the rainbow angle $\phi_r$
"observed" in the outgoing beam are defined with
$\sin \alpha=k_y/k$, and  $\sin \phi_r = k_y/k_{\perp}$.
We then obtain
\begin{equation}
\sin \alpha= \sin \phi_r \sin \theta = \sin \tilde{\phi}_r~\sin \theta~
\sqrt{1 + \frac{D}{E_{\perp}}}.
\label{eq:DeflAng}
\end{equation}
%
%
%\begin{equation}
%\sin \phi_r = k_y/k_{\perp}=\sin \tilde{\phi}_r~\sqrt{1+\frac{D}
%{E \sin^2 \theta}},
%\label{eq:rainAng}
%\end{equation}
%
It follows from Eq.~\eqref{eq:DeflAng}
%and Eq.~\eqref{eq:rainAng}
that for the incidence conditions such that
$E_\perp$ becomes
comparable with the depth of the attractive potential well $D$, the
refraction effects strongly alter the observed rainbow and maximum lateral
deflection angles. If the energy of the beam $E$ is fixed, decreasing
the grazing incidence angle $\theta$ leads to $\alpha$ that tends to
a constant non-zero value. This is in complete contrast with the $\alpha=0$
limit (pure specular reflection) that would be expected in the absence of the
projectile attraction to the surface. Similarly, if the angle $\theta$
is constant, decreasing the beam energy $E$ leads to increasing
deflection $\alpha$, whereas it would be constant without VdW attraction.

We first demonstrate the refractive effects by analyzing the elastic
intensity. % from the diffraction of fast Ne atoms on LiF(001), and we show that the experimental observations are inline with predictions of the simple model outlined above.
Fig.~\ref{fig:Charts}a displays the diffraction chart obtained by varying
the grazing incidence angle $\theta$, for a fixed energy $E$ of the incident beam. Here the radius of the Laue circle is proportional to $\theta$ while the angular distance $\Delta \alpha = G_y/k$ between diffraction orders stays constant. The reciprocal lattice vector of the LiF(001) surface is given by $G_y=2\sqrt{2} ~\pi/a$ where $a=4.03$~\AA~ is the LiF lattice constant.
The intensity of the diffraction orders oscillate with $k_\perp$, reflecting the interference of trajectories from the top and bottom of the corrugation function.
The general shape made of maxima aligned along nested V structures comes from the modulation of the intensity due to supernumerary rainbows.
The outermost structure corresponds to the classical rainbow angle~\cite{Schueller_2008} measured here as $\phi_r=31^{\circ}$ in good agreement with earlier results \cite{Gravielle_2011}.
At the lowest incidence angles $\theta$ (lowest $k_{\perp}$) the
$\alpha(\theta)$-dependence of the maximum deflection angle and thus of the rainbow angle
departs from linear, \textit{inline} with the refraction effect described by Eq.~\eqref{eq:DeflAng}.

Fig.~\ref{fig:Charts}c shows the diffraction chart obtained for different
experimental conditions. Now, the grazing incidence angle $\theta$ is fixed while the
total energy $E$ is varied. In this situation, the difference in Bragg angles
separating adjacent diffraction orders changes as $\Delta \alpha = G_y/\sqrt{2 M E}$
(see white dotted lines in Fig.\ref{fig:Charts}c). The maximum deflection angle $\alpha$
as well as the rainbow angle $\phi_r$ is constant for energies above
1.5 keV~\cite{Gravielle_2011}. At smaller energies $\alpha$ and so $\phi_r$ increases with
decreasing $E$. This effect is often interpreted as being due to a larger "effective" surface corrugation~\cite{Schuller_EPL_2008}. However, such an explanation does not hold here as follows from the calculated atom-surface interaction potential. Instead, the observed trends stem from the refraction of Ne atoms in the attractive VdW potential. This conclusion is supported by the comparison between the measured projectile energy dependence of the rainbow structure in the experimental diffraction chart,  and $\alpha(E)$-dependence calculated with Eq.~\eqref{eq:DeflAng}.

In order to obtain further insights into the Ne interaction with LiF(001) surface we have performed a dynamical diffraction theory study within the ASCA approximation. The
wave function of Ne projectile is given in the form $\Psi(y,z) = \sum_j \psi_j(z) e^{i j G_y}  \sqrt (2 \pi / G_y)$ leading to the set of close-coupling equations \cite{Manson1992} for diffraction orders j.
        %the
        \begin{equation}
            -\frac{1}{2 M} \left[\frac{\partial^{2}}{\partial z^{2}}
            - \left(j G_y\right)^2\right] \psi_j +
            \sum_{\ell} W_{j\ell} \psi_\ell = E_\perp \psi_j(z).
            \label{eq:cc}
        \end{equation}
Solving Eq.~\eqref{eq:cc} leads to the extraction of the scattering matrix \cite{Norcross1973} and thus of the intensities of the diffracted beams. The coupling matrix $W_{j\ell}$ is given by
        \begin{equation}
            W_{j\ell}(z) = \int_{-L/2}^{L/2} dx \int_{-L/2}^{L/2} dy~
            V(x,y,z)~ \frac{e^{i (\ell-j) G y}}{L}.
            \label{eq:w}
        \end{equation}
The Ne-surface interaction potential is searched in a form comprising
both planar and binary interaction terms
\begin{equation}
V(\rm{\bf{r}}) = - \mathcal{C} /z^3 + \sum_{n=1,2}
\frac{a_{n,s}}{\left|\rm{\bf{R}}_{s}-\rm{\bf{r}}\right|} e^{-\gamma_{n,s} \left|\rm{\bf{R}}_{s}-\rm{\bf{r}}\right|},
\label{eq:v}
\end{equation}
where $\rm{\bf{r}}=(x,y,z)$, and the sum runs over the $s=\rm{Li}, \rm{F}$ lattice sites located at $\rm{\bf{R}}_{s}$. For the planar attractive VdW interaction $- \mathcal{C}/z^3$\cite{Hoinkes1980}, the $z=0$ is set at the plane of surface F$^-$ ions. 
The second term in Eq.~\ref{eq:v} is a sum of screened coulomb binary interaction potentials. 
These forms decay exponentially at large distances and have proven their efficiency in atom-atom scattering in general \cite{zbl} and also in reproducing GIFAD data for various systems \cite{Schuller_EPL_2008,Schueller_2008}. 
%While $n=4$ was used in Ref.~\cite{Schuller_EPL_2008} at higher $E_{\perp}$ energies, we found that for the present scattering conditions using $n=2$ allows an excellent description of the data as follows from the results presented in Fig.~\ref{fig:Exp_Ib}.
%It is noteworthy that with an appropriate choice of  $a_{n,s}$, the attractive part of the potential can be described using binary interactions, as discussed below. 
%In overall, however, an explicit account for the planar attractive potential term improves the description of the experimental data. 
We have also accounted for the rumpling of $-0.05$ \AA \cite{Schuller2010} in position of surface Li atoms with respect to the LiF(001) surface plane.

Representation given by Eq.~(\ref{eq:v}) allows an analytical closed-form expression for the coupling matrix so that the diffraction chart can be calculated in seconds allowing on-the-fly optimization of the potential. 
%In practice we proceed as follows, 
%starting with a set of parameters $\mathcal{C},a_{n,s},\gamma_{n,s}$, equations Eq.~\eqref{eq:cc} are solved allowing to find the intensities of the diffracted beams as function of $E_{\perp}$.
Here the $\mathcal{C},a_{n,s},\gamma_{n,s}$ starting parameters were derived from the binary potentials deduced in \cite{Miraglia_2017} from the density functional theory DFT calculations of the Ne / LiF(001) interaction.
The intensities of the diffracted beams as function of $E_{\perp}$, derived by solving equations Eq.~\eqref{eq:cc} are compared with experimental values for the energies $E_\perp$, between 20 and 150 meV.
%The binary potentials deduced in \cite{Miraglia_2017} from the density functional theory DFT calculations of the Ne / LiF(001) interaction.
%Starting with a set of parameters $\mathcal{C},a_{n,s},\gamma_{n,s}$, the intensities of the diffracted beams as function of $E_{\perp}$ are derived by solving equations Eq.~\eqref{eq:cc}.
%are solved allowing to find the intensities of the diffracted beams as function of $E_{\perp}$. 
%The potential parameters are then optimized using minimization of the least squares deviation between calculated and measured intensity of the $j=0, \pm 1, ... \pm 9$ diffraction orders for the energy of the motion perpendicular to the surface, $E_\perp$, between 20 and 150 meV. 
The potential parameters are then optimized using a least squares minimization.

%to the measured intensity of the $j=0, \pm 1, ... \pm 9$ diffraction orders for the energies $E_\perp$, between 20 and 150 meV. 
%The optimization is initiated with parametrization of the binary potentials deduced in \cite{Miraglia_2017} from the density functional theory DFT calculations of the Ne / LiF(001) interaction and using only n=2 in Eq.~\ref{eq:v}. 
%We start an iterative procedure with parametrization of the binary potentials deduced in \cite{Miraglia_2017} from the density functional theory DFT calculations of the Ne / LiF(001) interaction. 
%We found that this choice allows faster convergence in our conditions as compared to the potential proposed by Sch\"{u}ller and collaborators \cite{Schuller_EPL_2008}. 
%The latter allows an excellent description of the rainbow scattering however at relatively high $E_{\perp}$ energies. 
%In this respect, potential found in our work can not be considered as a universal potential of the Ne / LiF(001) interaction. 
%Rather it is well suited to describe the atom/surface scattering for $E_\perp$ between 20 and 150 meV, and, most importantly, it accounts for the attractive physisorption well which is the focus of the present work.
%From the fit we obtain the following parameters: $a_{1,F}=74.7578,~\gamma_{1,F}=1.6108;~a_{2,F}=-1.4581,~\gamma_{2,F}=0.9940;~a_{1,Li}=23.0115,~\gamma_{1,Li} = 1.9288;~a_{2,Li}=-1.8785,~\gamma_{2,Li}= 1.214;~\alpha=3.6052 \times 10^{-3}$.
%The parameters derived from our fit are reported below under Ref~\cite{ZBLparameters}.
Figure~\ref{fig:Exp_Ib} shows an excellent agreement with the data over the complete energy range. 
The $\mathcal{C},a_{n,s},\gamma_{n,s}$ parameters extracted here are reported in Ref~\cite{ZBLparameters}.
%The comparison between experimental data and dynamical diffraction theory calculations is shown in Fig.~\ref{fig:Exp_Ib}. We obtain an excellent agreement over ten diffraction orders in the complete energy range.
Below 20 meV, the theory indicates numerous narrow peaks associated with bound state resonances. To avoid singular behavior, this energy range was not included in the fit.
%Here the large mass of neon is clearly a handicap, not only the density of bound states is much larger but the coupling between multiple diffraction peaks is obviously more complex than for helium where resonances were observed for only $j=0, \pm 1$ below 25 meV~\cite{Debiossac_PRL_2014}.

The planar potential $V(z)$ deduced from our fit is shown in Fig.~\ref{fig:Charts}b.
It is characterized by a $D=10.4$~meV attractive potential well, in good agreement with thermal atom diffraction data~\cite{Hoinkes1980,Boato76,Vidali91}. 
Note that a higher value D = 13.5 meV has been also reported ~\cite{Semerad_1987} based on observation and modeling  of bound state resonance but the difficult line assignment has never been confirmed.
%Note that a higher value $D=15$~meV has been also reported, but based on a theoretical estimation \cite{Hoinkes1980}.
With our parametrization, the attractive well is obtained owing to both the $\mathcal{C}/z^3$ term and the attractive part of the binary interaction potentials. 
To further test the robust character of our surface potential model, we performed another potential optimisation imposing $\mathcal{C}=0$. 
In this situation the attraction to the surface is entirely given by the negative terms ($a_{n}<0$) of the binary potentials. 
Our fit also reported in Ref~\cite{ZBLparameters} and displayed in Fig.~\ref{fig:Charts}b) then indicates a value $D=9.9$~meV, very close to $D=10.4$~meV reported above.
We have also tried to use the binary potentials derived in Ref~\cite{zbl,Schuller_EPL_2008} but the convergence is much slower in spite of having twice as many terms in the expansion (n=4) in Eq.~\ref{eq:v}.
%This highlights that none of the available potential is universal, the one in Ref~\cite{zbl} was derived from collisions of keV to MeV projectiles at quasi-normal incidence but found too repulsive in Ref~\cite{Schuller_EPL_2008,Schueller_2008} probing rainbow angles at grazing incidence with $E_\perp$ in the 0.1 eV-50 eV range .
This highlights that extrapolation is a difficult task, Z.B.L. interaction potentials~\cite{zbl} optimized to fit scattering of keV to MeV projectiles at quasi-normal incidence were found too repulsive when probing rainbow angles at grazing incidence with $E_\perp$ in the 0.1 eV-50 eV range~\cite{Schuller_EPL_2008,Schueller_2008}.
For $E_\perp$ between 20 and 150 meV, our study indicates that both Ref~\cite{zbl,Schuller_EPL_2008} underestimate the attractive potential responsible for physisorption well, which is the main focus of this work. 

    %
    % FIGURE 4---------------
    %
    \begin{figure}
        \includegraphics[width=0.80\linewidth,draft = false] {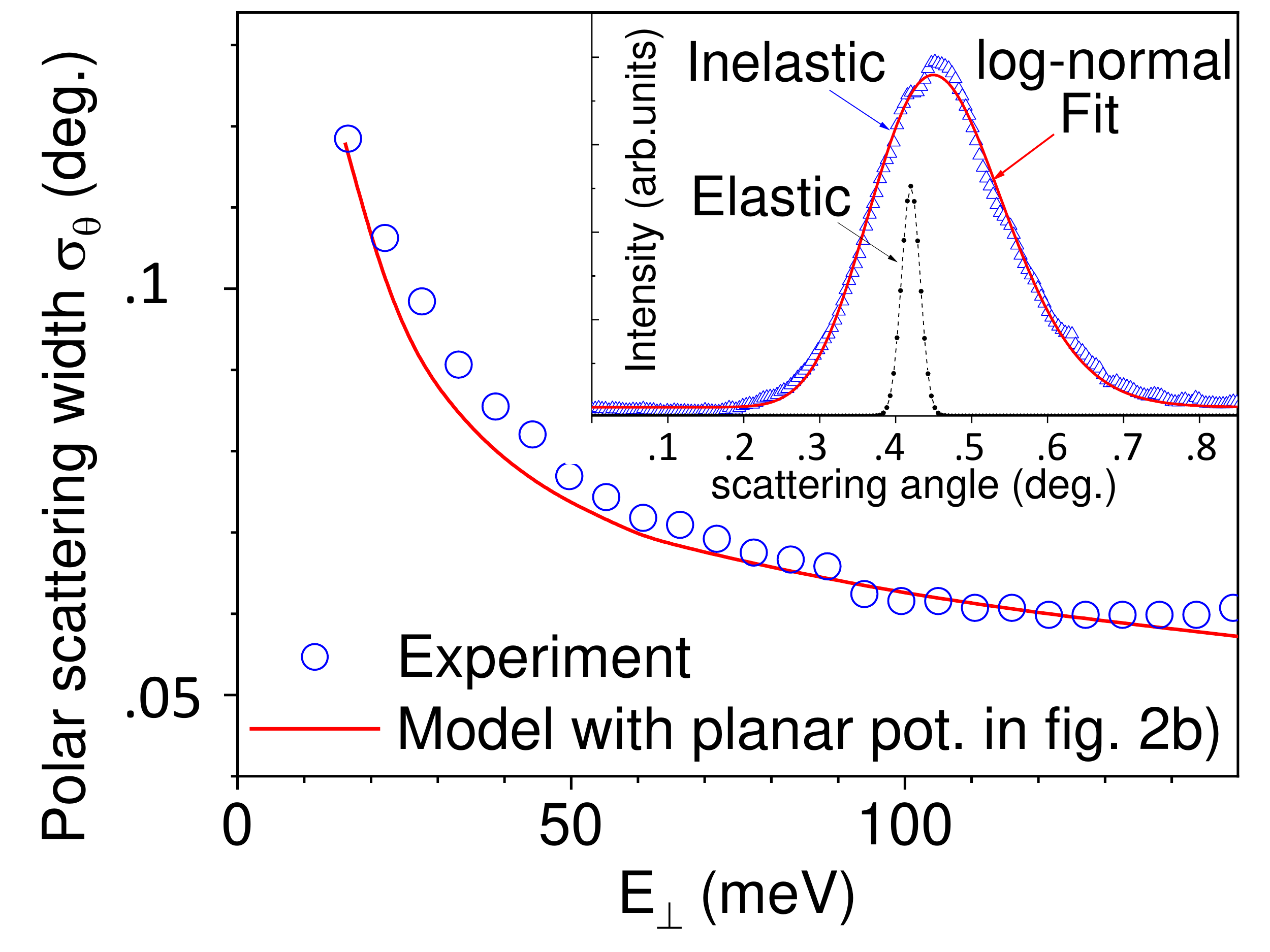}
        \caption{The width $\sigma_\theta$ of the polar inelastic scattering profile. Dots (line) show the experimental (model) results as a function of the energy of the motion in the $(y,z)$-plane, $E_\perp$. The inelastic profile is obtained from experimental data by subtracting the elastic contribution. $\sigma_\theta$ is extracted from the inelastic profile using a log-normal fit as shown in the insert for 500 eV neon at 0.42 deg.
            %    It is extracted from the log-normal fit where the scale parameter $w$ is %$\sigma=2\theta\sqrt{e^{w^2} (e^{w^2} -1)} $.
            \label{fig:sigma_theta}}
    \end{figure}

Another evidence for refraction of Ne atoms is provided by the analysis of the \emph{inelastic} polar scattering profile. It is obtained from the experimental
data by subtracting the sharp peaks of the elastic contribution on the Laue circle \cite{Debiossac_Nim_2016} (see \textit{e.g.} Fig.~\ref{fig:y13_h221_2D}).
The inset in Fig.\ref{fig:sigma_theta} shows that the resulting intensity
distribution over polar exit angles $P(\theta_{out})$ is well fitted by a log-normal profile \cite{Debiossac_Nim_2016,Roncin_PRB_2017} allowing its width $\sigma_\theta$ to be define.
Fig.~\ref{fig:sigma_theta} shows a clear increase of $\sigma_\theta$ with decreasing $E_{\perp}$. This is despite the Ne atoms interacting less strongly with the surface.
This apparent contradiction can be understood from the simple scattering model outlined at the beginning.
For simplicity we consider only the scattering in the specular plane ($x,z$).
In our experimental conditions the distance of closest approach to the surface amounts to few \AA and is much larger than the amplitude of thermal motion of the surface atoms $\sigma$.
A perturbative approach then predicts a log-normal scattering profile with a relative angular width  $\sigma_{\tilde{\theta}}/\tilde{\theta}\propto\Gamma\sigma$
\cite{Manson_PRB_2008,Roncin_PRB_2017}. Here $\Gamma$ is the logarithmic
derivative of the interaction potential between the Ne atom and the surface
evaluated at the turning point of the trajectory.
The "effective" polar scattering angle $\tilde{\theta}$ and the "observed" polar scattering angle $\theta$ are related as $\tilde{\theta}^2=\sqrt{\theta^2+D/E}$ so that, for small angular broadening we obtain %$\sqrt{1+D/E_{\perp} }
$d\tilde{\theta} = d\theta/\sqrt{1+D/E_{\perp} }$.
%We assume that the proportionality approximately holds for finite $\sigma_{\tilde{\theta}}$ and $\sigma_{\theta}$.
While $\sigma_{\tilde{\theta}}$ is fixed by the scattering conditions in the
vicinity of the surface, the observed polar angle width $\sigma_{\theta}$
increases with decreasing $E_{\perp}$.
The proportionality factor between $\sigma_{\theta}$ and $\Gamma$ is adjusted at $E_\perp$= 120 meV and the model displayed in Fig.~\ref{fig:sigma_theta} reproduces qualitatively well the drastic increase at low energy.

In conclusion, we reported well resolved elastic diffraction of neon atoms from LiF(001) surface.
Combined with a fast algorithm using dynamical diffraction theory, the large
number of diffraction orders observed experimentally with these heavy projectiles
allows a refinement of the model projectile-surface interaction potential.
The quantitative agreement reached between the theory and experiment is remarkable
and its sensitivity to the potential parameters yields a robust estimate of
the depth of the attractive van der Waals potential well.
Our work also sheds light on the analogies between geometrical optics and projectile scattering at surfaces, where the refraction of atomic beams manifests itself via the significant increase of the elastic azimuthal rainbow angle as well as via the drastic increase of the polar inelastic scattering profile.
Even in the absence of observation of any bound states resonances, the accurate determination of the diffracted intensities at grazing incidence conditions is a powerful strategy to measure the van der Waals forces needed to refine \textit{ab initio} theoretical descriptions. In this energy range, neon atoms offer an increased sensitivity owing to their shorter wavelength and even inelastic effects contribute to an accurate self-consistent determination of attractive forces. Further investigations are needed to develop a quantitative theory for inelastic diffraction of fast atoms supporting our interpretation of the inelastic scattering profiles.

\section{Acknowledgment}
J.E. Miraglia and M.S. Gravielle are kindly acknowledged for providing a compact form of their results \cite{Miraglia_2017}.

\bibliography{Neon}  % nom du fichier .bib  ici Neon.bib

\end{document}